Fill and dump measurement of the neutron lifetime using an asymmetric magneto-gravitational trap

C. Cude-Woods,[1, 2] F. M. Gonzalez,[3, 4, 5] E. M. Fries,[6] T. Bailey,[1] M. Blatnik,[6] N. B. Callahan,[7]
J. H. Choi,[1,2] S. M. Clayton,[8] S. A. Currie,[8] M. Dawid,[3,4] B. W. Filippone,[6] W. Fox,[3,4]
P. Geltenbort,[9] E. George,[10] L. Hayen,[1,2] K. P. Hickerson,[6] M. A. Hoffbauer,[8] K. Hoffman,[10] A. T. Holley,[10]
T. M. Ito,[8] A. Komives,[11] C.-Y. Liu,[1,2] M. Makela,[8] C. L. Morris,[8] R. Musedinovic,[1,2] C. O'Shaughnessy,[8]
R. W. Pattie, Jr.,[12] J. Ramsey,[5] D. J. Salvat,[3,4] A. Saunders,[8, 5] E. I. Sharapov,[13] S. Slutsky,[6] V. Su,[6] X. Sun,[6]
C. Swank,[6] Z. Tang,[8] W. Uhrich,[8] J. Vanderwerp,[3,4] P. Walstrom,[8] Z. Wang,[8] W. Wei,[6] and A. R. Young[1, 2]

[1]Department of Physics, North Carolina State University, Raleigh, NC 27695, USA
[2]Triangle Universities Nuclear Laboratory, Durham, NC 27708, USA
[3]Department of Physics, Indiana University, Bloomington, IN, 47405, USA
[4]Center for Exploration of Energy and Matter, Indiana University, Bloomington, IN, 47405, USA
[5]Oak Ridge National Laboratory, Oak Ridge, TN 37831, USA
[6]Kellogg Radiation Laboratory, California Institute of Technology, Pasadena, CA 91125, USA
[7]Argonne National Laboratory, Lemont, IL 60439, USA
[8]Los Alamos National Laboratory, Los Alamos, NM, USA, 87545, USA
[9]Institut Laue-Langevin, CS 20156, 38042 Grenoble Cedex 9, France
[10]Tennessee Technological University, Cookeville, TN 38505, USA
[11]DePauw University, Greencastle, IN 46135, USA
[12]East Tennessee State University, Johnson City, TN 37614, USA
[13]Joint Institute for Nuclear Research, 141980 Dubna, Russia

**Abstract** The past two decades have yielded several new measurements and reanalyses of older measurements of the neutron lifetime. These have led to a 4.4 standard deviation discrepancy between the most precise measurements of the neutron decay rate producing protons in cold neutron beams and the lifetime measured in neutron storage experiments. Measurements using different techniques are important for investigating whether there are unidentified systematic effects in any of the measurements. In this paper we report a new measurement using the Los Alamos asymmetric magneto-gravitational trap where the surviving neutrons are counted external to the trap using the fill and dump method. The new measurement gives a free neutron lifetime of $\tau_n = 877.1(2.6)_{\text{stat}}(0.8)_{\text{syst}}$. Although this measurement is not as precise, it is in statistical agreement with previous results using *in situ* counting in the same apparatus.

## Introduction

The unitarity of the Cabibbo-Kobayashi-Maskawa (CKM) matrix, which describes the weak mixing between the six flavors of quarks, provides a stringent test of the standard model of particle physics. The first row is most easily measured because the dominant matrix element, $V_{ud}$, can be inferred from nuclear (including neutron) beta decay measurements. Neutron decay has the potential to provide $V_{ud}$ to a precision exceeding that achieved in nuclear beta decay because of smaller uncertainties due to the simpler structure of the neutron [1-5]. However, the current data set[6-18] shows a large discrepancy in lifetimes obtained by measuring the rate of neutron decay resulting in protons in the final state[12] and neutron lifetimes measured by counting surviving neutrons stored in a trap experiments.[19] An independent analysis of systematic uncertainties in the beam experiment[20] suggests that charge exchange on residual gas was not sufficiently analyzed in the beam experiment and that this effect might explain the discrepancy, but this needs to be tested. Measurements using different techniques that might identify systematic uncertainties that have not been identified are necessary to confirm or eliminate this discrepancy.

In this report we present new data taken using the asymmetric magneto-gravitational ultra-cold neutron (UCN) trap[21] that was used for recently reported lifetime measurements[14, 16, 18] but using a different counting technique. Rather than *in sit*u "dagger" detector counting we have unloaded the neutrons and counted them external to the trap, the so-called fill and dump technique.

## Experiment

The Los Alamos Ultracold Neutron Facility (LAUNF) produces a high density of ultracold neutrons (UCN) by down scattering spallation neutrons produced using pulsed 800 MeV protons from the Los Alamos Neutron Science Center (LANSCE) accelerator at the Los Alamos National Laboratory (LANL) incident on a solid deuterium converter.[22, 23] These neutrons are transported using nickel phosphorous coated guides[24, 25] through a 6 T pre-polarizing superconducting solenoid magnet that selects high field seeking UCN and an adiabatic spin flipper that converts them to low field seeking UCN.

The low field seeking neutrons are loaded into the magnetic trap by removing a small section of the Halbach array, the "trap door". The neutrons are stored in the trap by replacing the "trap door". In the previous work,[18] those stored UCN were counted in place by lowering a detector into the trap. This provided a short counting time of a few seconds and minimized errors introduced by the coupling between the phase space of the UCN and the time at which the UCN are counted.

Here we report a set of measurements made using the same experimental apparatus with the same UCN loading procedures, but with the neutrons counted at the end of storage by unloading through the trap door into an external detector. The time constant for counting UCN in a dagger with a 20 nm coating lowered to the bottom of the trap is 7.1(2) s. The time constant for unloading UCN in this experiment is 26.8(8) s. Although these experiments share many sources of systematic uncertainties, the coupling between phase space evolution and counting could be much larger in this experiment. These results provide a test of the method used to estimate the size of this systematic uncertainty in the previous experiment.

A layout of the experiment is shown in Figure 1. UCN are produced by 0.5 s long pulses of ~0.1 mA of proton beam delivered every 5 s. UCN are produced in a shielded volume of solid deuterium $(\mathrm{SD}_2)$ and are piped out of the shielding crypt with a 1m rise and a 4 m long chicane. The source vacuum that contains the SD$_2$ is isolated from the experiment by a 50 μm aluminum foil at the center of the 6 T super conducting solenoid polarizing magnet. This allows a vacuum of order $10^{-7}$ torr to be maintained in the UCN trap despite $10^{-4}$ torr pressure spikes in the source due to beam heating of the source.

In Figure 1 UCN enter from the left. The 6 T magnetic field in the polarizing magnet selects only high field seeking UCN. UCN then pass through the adiabatic fast passage spin flipper[26] where they are converted into low field seeking UCN. A large buffer volume serves to filter out fluctuations in the beam intensity due to the proton beam delivery, and an active cleaner removes and counts neutrons with velocities too high to be trapped. The active cleaner and monitor in the buffer volume are used to normalize the experiment.

The trap[21] is formed of vertically oriented toroids, one with a minor radius of 1.0 m and a major radius of 0.5m, the other with a major radius of 0.5 m and a minor radius of 1.0 m. These mated on axis where the surface of both toroids had a radius of 1.5 m The torii are cut off at 50 cm above their lowest point and are contained in a vacuum jacket. The surface of the toroids was made from rows of neodymium boron iron permanent magnets arranged as a Halbach array.   The minimum field at the surface of the toroids was about 0.8 T, providing a trapping potential for low field seeking UCN of 48 neV. This trapping potential is approximately the same as the kinetic energy a UCN attains when dropped at rest from a height of 50 cm. UCN that are fully repelled by the 50 neV trapping potential over the entire trap surface cannot and bounce above the upper edge of the trap

A set of copper coils arranged normal to the surface of the permanent magnet torus provided a field that eliminated field zeros and ensured the depolarization rate of stored UCN was negligible.  The holding fields was about 60-120 Gauss, depending on position in the trap.

A cleaner whose surface was a combination of polyethylene and $^{10}$B/ZnS was lowered to a level of 38 cm above the bottom of the trap to remove high energy UCN, well below the upper edge of the trap. This ensured that the remaining population was trapped.  The fraction of untrapped UCN after cleaning was previously measured and found to be small.[18]

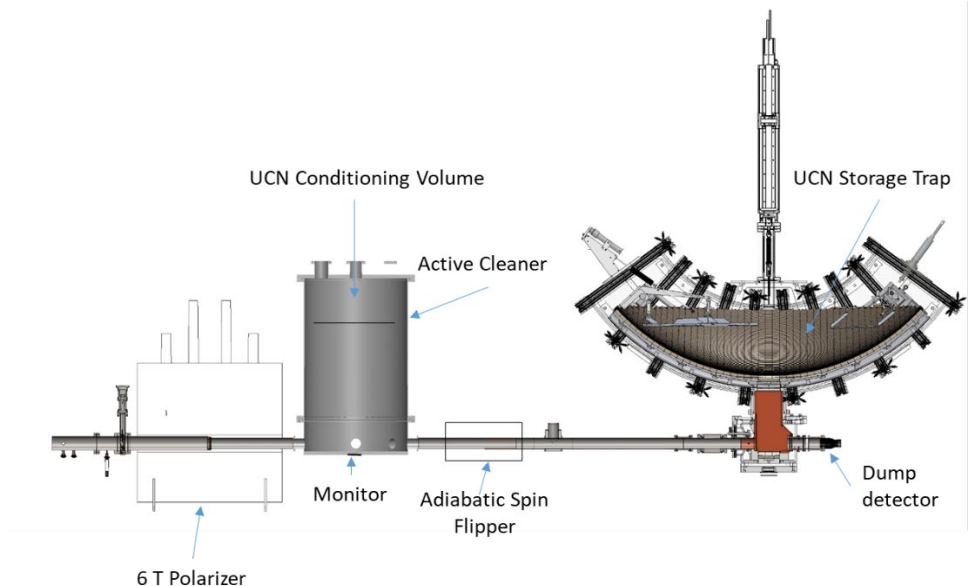

Figure 1. Experimental layout showing the location of the UCN detectors used in this experiment. The source is off the picture to the left. The dump detector is the primary counter for this experiment.

The sequence of loading, storing and counting the UCN is shown in Figure 2. The sequence was controlled by three valves, the flapper, butterfly, and trap door. The trap door is a section of magnets that can be lowered to open a hole in the bottom of the trap. The flapper had three positions: an up position, used to raise and lower the trap door; a 45˚ position, to optimize transport into the trap; and a down position, to minimize UCN interactions with the lowered trap door when unloading the trapped UCN. The butterfly valve located upstream of the loading port, was open for loading and closed for counting. The trap door was lowered for loading and counting and raised to store UCN in the trap.

The UCN detector was comprised of a 76 mm diameter photomultiplier tube viewing a $^{10}$B/ZnS detector.[27] The $^{10}$B/ZnS was inside of the vacuum and the light was transported to the photomultiplier via a Lucite light pipe that also served as the vacuum window. The signals were processed using a timing filter amplifier, which integrated for 100 ns, and a single channel analyzer (MSC4 from FAST ComTec Communication Technology GmbH). These were then counted in a multi-channel multi-scalar that also recorded the monitor signals as a function of time.

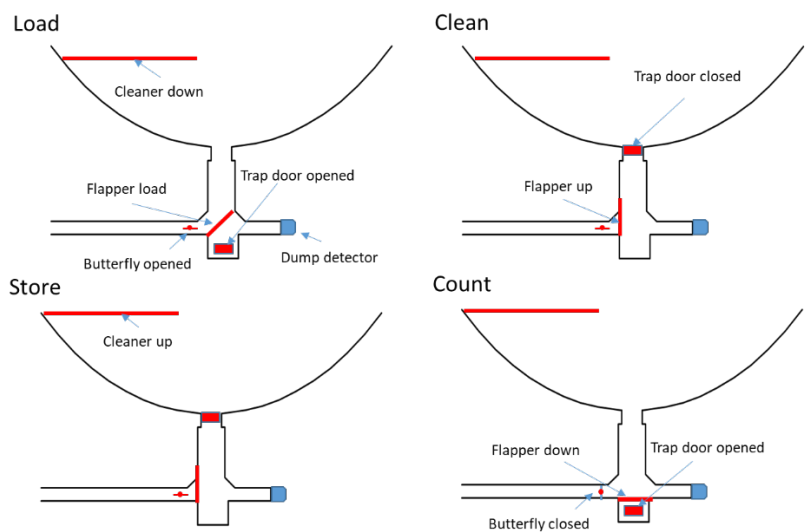

Figure 2. schematic showing the valve configurations for loading, cleaning, storing, and counting. The valve changes for each step are labeled.

### Analysis

The results presented here were obtained from 189 runs taken over a week of running. Data were taken in run sequences (octets) that included holding times of 20, 1550, 1550, 50, 100, 1550, 1550 and 200 s, a sequence designed to cancel normalization drifts. A spectrum from a 20 s holding time run is shown in Figure 3, where counting for fill and dump is compared to *in situ* counting.

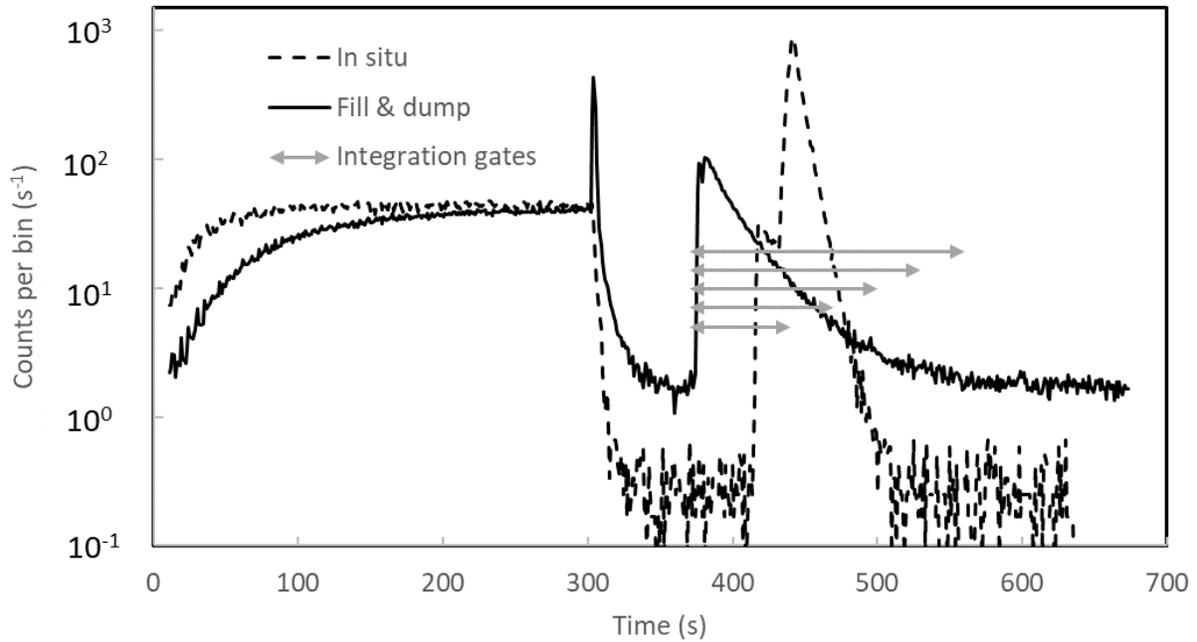

Figure 3. Counting spectrum in fill and dump mode (solid) and *in situ* counting (dashed). The unloading times are 28 s and 7 s for fill and dump and in situ counting respectively. The integration gates for fill-and-dump are shown as the solid gray double arrows. The double peak in the *in situ* counting curve is because UCN were counting at two different dagger positions. The loading curves are different because of the different counter positions.

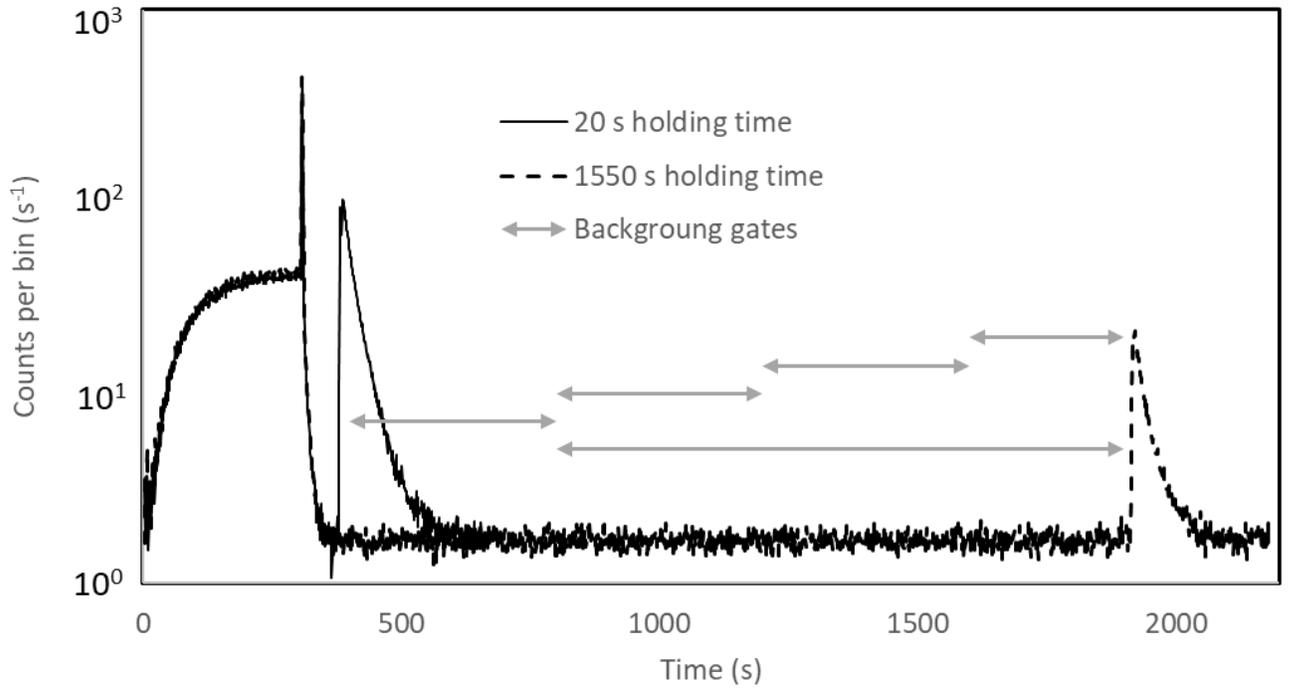

Figure 4. The figure shows a comparison of the average of the short holding time runs with the long holding time runs. The gray double arrows show the background gates that have been used in the analysis.

Yields we calculated for each run, k, as

$$Y_{0k} = \frac{S_k - B_k}{N_k}$$

$$\Delta Y_{0k} = \frac{\sqrt{S_k + B_k}}{N_k} \quad (1)$$

$$N_k = \sum_i a_i \text{RHAC}_i$$

where $S_k$ is the sum of counts in a 130 s long counting gate beginning at the time the trap door is opened, $B_k$ is the background obtained from a nearby long holding time run (see Figure 3). The counting rate from the round house active cleaner ($\text{RHAC}_i$) was binned into 25 s long groups. These were summed with weights for each group, $a_i$, that were optimized to reduce the sensitivity of the yields to beam fluctuations. The weights were fixed at values obtained by fitting a larger independent data set taken with in situ counting.

The background counting rate was obtained from the long holding time runs, which had a long background region from the end of cleaning to the beginning of counting. Backgrounds for the short holding time runs were obtained from nearby long holding time runs. This background analysis was performed after unblinding the data because of concerns about the lack of a background region in the short holding time runs. It resulted in a shift in the lifetime 1.2 s.

Variations in the width of the foreground integration gate and the location of the background region (shown in Figures 3 and 4) were studied to establish a system systematic error associated with the background subtraction The lifetime was taken as an average over these tests (shown in Figure 5) and a systematic uncertainty due to the background model was taken as the standard deviation of these values.

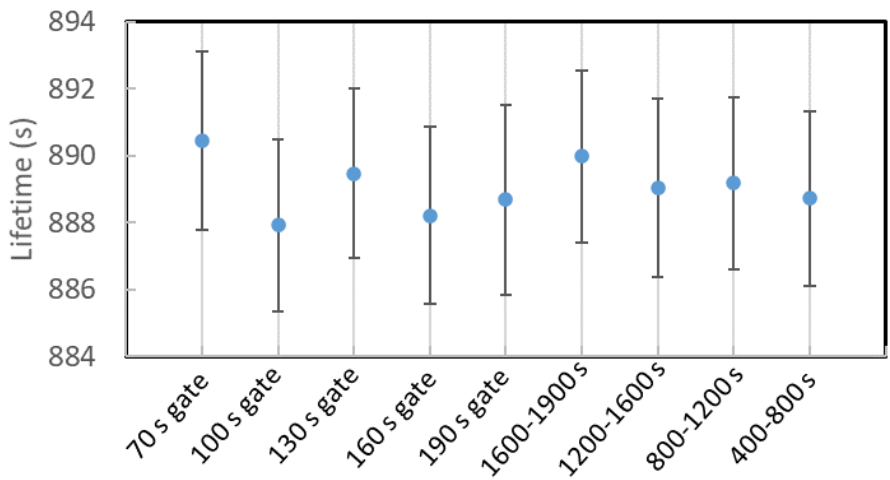

Figure 5. Data points are the results of the background systematic study. The first 5 points, labeled with the foreground gate width, used a background region of 400-1600 s. In the last four points the background region was varied with the foreground gate width fixed at 130 s.

This normalization is based on the higher velocity part of the spectrum with energies above the trapping potential. Degradation of the surface of the solid deuterium of the source results in both reduced output and hardening of the UCN spectra.[28] These time dependent normalization changes were accounted for by using a linear correction based on the ratio of the round house monitor (RHMON) to the round house active cleaner (RHAC) detector (see Figure 1).

The nonzero corrections were applied to the reported yields.

$$Y_k = \frac{Y_{0k}}{1-b(r_k - \overline{r}_k)}$$

$$\Delta Y_k = \frac{\Delta Y_{0k}}{1-b(r_k - \overline{r}_k)} \quad (2)$$

$$r_k = \frac{RHMON}{RHAC}$$

The constant $b$ was obtained fitting the same independent data set used to get the $a_i$. The neutron lifetime is introduced using a correction factor ($CF$) calculated using the neutron lifetime, $\tau$, and the holding time, $t_k$:

$$Y_{calc,k} = e^{-\frac{t_k}{\tau}}$$

$$R_k = \frac{Y_k}{Y_{calc,k}} \quad . \quad (3)$$

$$CF = \overline{R}_k$$

The UCN source output changes due to the exact geometry of the solid deuterium source. To keep the source output high, the deuterium crystal must be reformed approximately every 2-3 days. The average normalization factor $CF$ is taken over surrounding short runs (holding times of 20s, 50s, 100s, and 200s), with discrete breaks to account for these source rebuilds.

The counting uncertainties, $\Delta Y_k$, were assumed to have a Gaussian distribution. The neutron lifetime is obtained by fitting $\tau$ to minimize a chi square, $X^2$:

$$X^2 = \sum_k \frac{(Y_k - Y_{calc,k} CF_k)^2 DQE}{(\Delta Y_k)^2}, \quad (4)$$

where $DQE = 0.82$ is a factor to account for fluctuations not captured by the counting uncertainties. Without this factor the reduced $X^2$ was 1.2. This is consistent precious observation.[18] The counting uncertainties were assumed to have a Poisson distribution. Here, the number of degrees of freedom, *df*, is one less than the number of yields. The statistical uncertainty in the lifetime is taken as the change in $\tau$ needed to increase the $X^2$ by 1.

The yields measured here, in fill and dump mode, were only 32% of those measured in *in situ* counting. Transport into and out of the trap is known to be lossy. This is the result of gaps in the guides leading into the trap and depolarization in the complicated field structure with the trap door removed, which we speculate leads to spin-flip and neutron loss as neutrons move in and out of the trap.

There is a bias in the lifetime due to the assumption of Poisson statistics. The bias arises because larger counts have higher weights in a calculated mean. This effect is larger at longer times because of the lower average count and higher variance relative to shorter holding times. This correction is larger here than in Ref.[18] because of the lower UCN counts. This biases the fitted lifetime to be longer than the actual lifetime. Three different methods were used to account for this bias in the high precision measurements.[18] Here, the bias was estimated by Monte Carlo. A large number of data sets was generated and fitted to obtain lifetimes with a known parent lifetime as a function of the number of initial neutrons. The difference between the fitted and actual lifetime was the statistical bias. This was tabulated as a function of initial count and used to calculate the correction on a run by run basis. The net correction is -0.71 s with negligible uncertainty.

The nominal holding times were used in the fit. The mean counting time can be affected by evolution of the UCN phase space during storage in the trap because different regions of phase space can have different unloading times. The difference in the change of the actual and nominal holding times between long and short holding time runs was found to be 0.09±0.16 s, indicating a negligible effect from phase space evolution in the trap.

The pressure in the trap was monitored with a cold cathode vacuum gauge. A correction to the lifetime was computed based on these pressures averaged across the data set and the cross sections reported in references [29, 30]. The residual gas was assumed to be dominated by water; an assumption supported by previous mass spectrometer measurements.

A summary of the systematic uncertainties is given in table 1.

| Effect | Correction | Uncertainty |
|---|---|---|
| Uncleaned | -- | 0.11 |
| Heated | -- | 0.08 |
| Residual gas scattering | 0.18 | 0.10 |
| Depolarization | -- | 0.07 |
| Dead time correction | -0.02 | 0.02 |
| Phase space evolution | -0.17 | 0.15 |
| Background model | -- | 0.81 |
| Uncorrelated sum | 0.84 | |

Table 1. Systematic uncertainties and corrections to the neutron lifetime are given in seconds. The Uncleaned, Heated, and Depolarization systematic uncertainties were taken from Ref. [18].

Some of the correction (deadtime, phase space evolution and residual gas correction) and the background analysis were applied after unblinding the data. The average yields as a function of holding time along with the lifetime fit is shown in Figure 4. We find $\tau_n = 877.1(2.6)_{\text{stat}}(0.8)_{\text{syst}}$ s, where the first uncertainty is statistical and the second is systematic. Figure 5 compares the most precise beam (Yue)[12] and trap lifetime measurements (Gonzalez)[18] with the current result. A $X^2$ test gives the probability (assuming Gaussian statistics) of the previous beam and trap measurement being consistent of 1.2×10$^{-5}$, the current result and the beam result of 2×10$^{-3}$, and the current result and the trap result of 0.7. This measurement does not identify any discrepancy in the Gonzalez *et al.*[18] measurement that can help to resolve the difference between the beam and trap results.

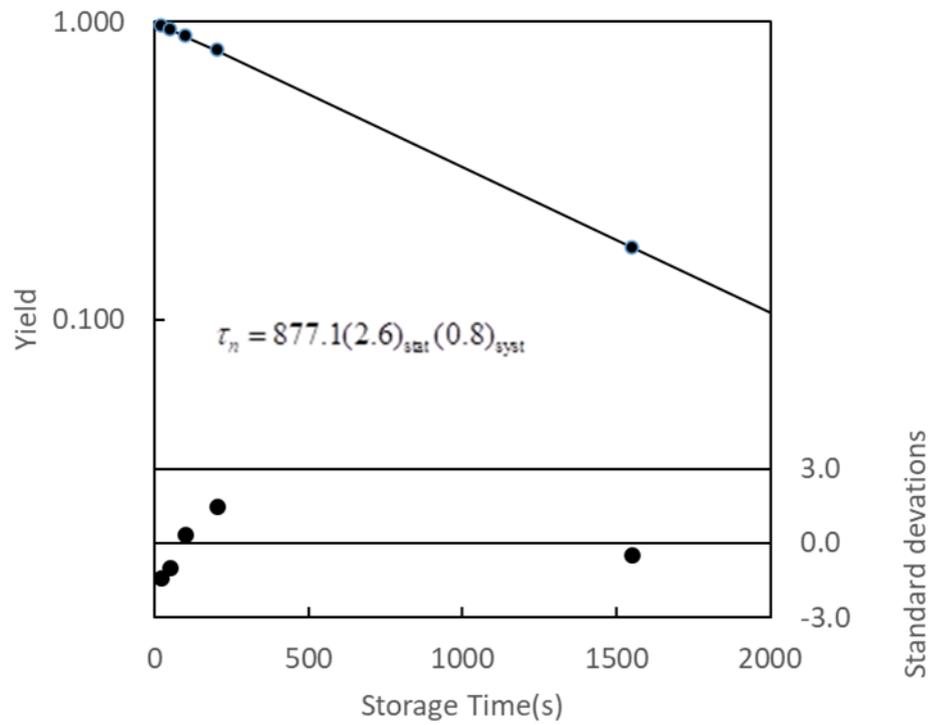

Figure 6. Plot of average yield vs time, lifetime fit, and residuals.

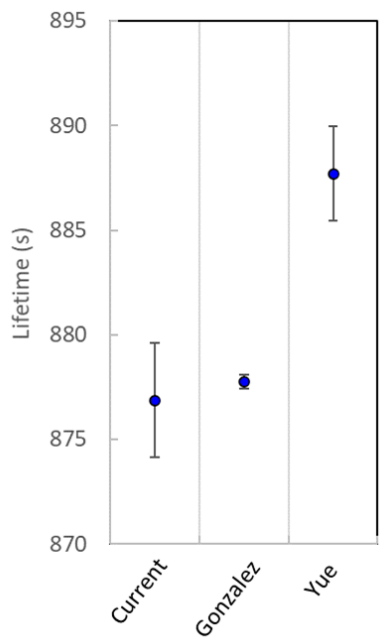

Figure 7. Comparison of the most precise beam[18] and trap lifetimes[12] along with the current result. The plotted uncertainties are the quadratic sum of the statistical and systematic uncertainties for each measurement.

## Conclusions

The neutron lifetime in the Los Alamos magneto-gravitational trap measured using the fill and dump method agree with those using *in situ* counting in the same trap. This very different counting method has many disadvantages when compared to in situ counting but is potentially sensitive to unidentified systematic uncertainties in in-situ counting. However, we have not identified any new systematic errors that can explain the 4.4 standard deviation difference between the beam and trap results associated with the counting method.

## Acknowledgments

This work is supported by the LANL LDRD program; the U.S. Department of Energy, Office of Science, Office of Nuclear Physics under Awards No. DE-FG02-ER41042, No. DE-AC52-06NA25396, No. DE-AC05-00OR2272, and No. 89233218CNA000001 under proposal LANLEDM; NSF Grants No. 1614545, No. 1914133, No. 1506459, No. 1553861, No. 1812340, No. 1714461, No. 2110898, and No. 1913789; and NIST precision measurements grant.